\documentclass[vecphys]{svmult}

\usepackage{makeidx}         
\usepackage{graphicx}        
\usepackage{psfig}           
\usepackage{multicol}        
\usepackage{url}             
\usepackage[bottom]{footmisc}

\makeindex             

\begin{document}

\title*{Implications\,\, of\,\, H.E.S.S.\,\, observations\,\, of\,\, pulsar\,\, wind\,\, nebulae}


\author{O.C. de Jager\inst{1},\,\,A. Djannati-Ata\"i\inst{2}}


\institute{
Unit for Space Physics, North-West University, Potchefstroom 2520, South Africa
\and CNRS, Universite Paris 7, Denis Diderot, F-75205, 75005 Paris, France
}

\maketitle

\begin{abstract}
In this review paper on pulsar wind nebulae (PWN) we discuss the properties of such nebulae
within the context of containment against cross-field diffusion
(versus normal advection), the effect of reverse shocks on the evolution of offset ``Vela-like''
PWN, constraints on maximum particle energetics, magnetic field strength estimates based on spectral and spatial 
properties, and the implication of such field estimates on the composition of the wind. A significant
part of the discussion is based on the High Energy Stereoscopic System ({\it H.E.S.S.} or {\it HESS}) detection of the two evolved pulsar wind
nebulae Vela X (cocoon) and HESS\,J1825-137. In the case of Vela X (cocoon) we also review evidence of
a hadronic versus a leptonic interpretation, showing that a leptonic interpretation is favored for the
{\it HESS} signal. The constraints discussed in this review paper sets a general framework
for the interpretation of a number of offset, filled-center nebulae seen by {\it HESS}. These sources are found along the galactic
plane with galactic latitudes $|b|\sim 0$, where significant amounts of molecular gas is found.
In these regions, we find that the interstellar medium is inhomogeneous, which has an effect on the
morphology of supernova shock expansion. One consequence of this effect is the formation of offset pulsar wind nebulae
as observed.
\end{abstract}

\section{Introduction \label{sec:1}}
Even before the discovery of pulsars, pulsar wind nebulae (PWN)
like the Crab Nebula were identified as belonging to a class of
cosmic radio sources with relativistic electrons moving in magnetized plasmas
to give the continuum radiation as observed. Visionaries
like \cite{Gou65} already predicted that we should be able
to measure the magnetic field strength in PWN using the combination
of synchrotron and inverse Compton (IC) radiation. Following this, \cite{Ken84} were the first 
to provide us with a sophisticated one dimensional (1D) 
magneto hydrodynamical models (MHD) model of the Crab Nebula, which predicted a magnetic field strength distribution, consistent
with broadband multiwavelength (radio through very high energy gamma-ray) constraints \cite{dej92,Ato96,Hil98}.

The discovery of the Crab pulsar in 1968 confirmed suspicions that a rapidly spinning
neutron star should provide the energy input into the Crab Nebula, but soon questions
concerning the spindown of pulsars in relation to the evolution of the nebulae arose.
Whereas a few Crab-like remnants were discovered, Vela X, assumed to be
associated with the 11,000 year old Vela pulsar, raised the question about
the evolution of PWN as described by \cite{Wei80}.
More serious evolutionary studies of PWN in supernova remnants (SNR) were launched by \cite{Rey84a} and 
\cite{Rey84b}, but the offset of Vela X relative to the Vela pulsar raised
the question if Vela X is indeed associated with the Vela pulsar. We also focus on
Vela X in this discussion for the very specific reason that it serves as a prototype
of evolved PWN. 
A wealth of new information on Vela X and similar evolved PWN became recently available
as discussed in this review paper.

An excellent review of the structure and evolution of PWN from a radio and X-ray perspective
was recently given by \cite{Gae06}, whereas investigations on the 
population of galactic very high energy (VHE) H.E.S.S. sources considered to be associated with PWN
were given by \cite{Lem06} and \cite{Gal07}.

\section{The evolving definition of pulsar wind nebulae (PWN)\label{sec:evol}}
The term ``pulsar wind nebulae'' (PWN), or ``plerions'' (derived from
Ancient Greek {\it pleres}) -- a term coined by 
\cite{Wei78}, is described by the following:

\begin{itemize}
\item Filled center or blob-like form;

\item A flat radio spectrum where the radio energy flux depends on frequency $\nu$ as $F_\nu\propto \nu^\alpha$, with $\alpha\sim 0$ to -0.3;

\item A well-organised internal magnetic field with high integrated linear polarisation at high radio frequencies.

\end{itemize}

The origins of the last property will be revisited in Section \ref{vb}.
 
The main body of information on PWN came mainly from radio and X-ray observations, whereas 
interstellar absorption makes it difficult to detect most of these diffuse sources in optical as well.
The study of PWN was therefore confined to a study of the synchrotron component only,
which depends on the nebular field strength. This, and the fact that only radio and X-ray
observations (representing widely divergent particle populations) have been used in theoretical
studies, must have restricted progress in the understanding of PWN evolution and the
conversion of spindown power into energetic particles in such nebulae.

With the growth of X-ray Astronomy, aided by sensitive instruments like {\it Chandra} and {\it XMM-Newton},
the definition of PWN has been broadened. We may add the following three important aspects

\begin{itemize}
\item A torus and jet near the pulsar, with the direction of the jet reflecting the 
direction of the pulsar spin axis and the torus showing an underluminous region
inside a characteristic scale radius $R_s\sim 10^{17}$cm to $\sim 10^{18}$cm, 
believed to be the pulsar wind shock radius (see \cite{Ng04} for parameter fits
to such torii);

\item Evidence for re-acceleration of particles somewhere between the pulsar light cylinder
and $R_s$, leading to a hard X-ray spectrum with a photon index $\sim 1.5$ to $\sim 2$ near $R_s$
(a review thereof will be given Section \ref{sec:3});

\item Evidence for synchrotron cooling (spectral steepening) at $R>R_s$, with the
size of the PWN decreasing towards increasing energies, as seen from the Crab and several
other PWN. The photon indices of the cooled spectra range between 2.0 and 2.5
(see also Section \ref{sec:3}).

\end{itemize}

The drawback of having only radio and X-ray synchrotron information on PWN
was realised by \cite{dej95,dup95,dej96b}, who were probably the first to predict that PWN evolved beyond the Crab phase (i.e. those
would have field strengths smaller than the Crab) would accumulate sufficient amounts
of relic electrons so that the inverse Compton (IC) scattering of these PWN electrons on the 
CMBR and possibly far infrared photons from galactic dust would be detectable by space and 
ground-based $\gamma$-ray telescopes. Furthermore, current generation 
ground based $\gamma$-ray telescopes have an angular resolution of a few arcminutes \cite{Hin04},
so that the ratio $\theta_{\rm PWN}/\theta_{\gamma}$ of the PWN angular radius $\theta_{\rm PWN}$ 
relative to the telescope angular resolution $\theta_{\gamma}$ increases to values greater than unity
as the pulsar/PWN evolves beyond the Crab phase. They effectively become resolved, which
allows us to measure spatially resolved spectra to the resolution of the VHE $\gamma$-ray telescopes.
These complementary $\gamma$-ray observations then allow us to probe the electron spectra in the 
sources, as well as the associated magnetic field distribution, provided that comparable
electron energies contribute to the observed (spatially resolved) synchrotron and IC spectra.

The H.E.S.S. telescope has the advantage of a large field-of-view, few arc\-minute angular resolution
and good sensitivity against background rejection to probe extended sources \cite{Hin04}
which allowed this telescope to see for the first time a population of evolved (resolved) VHE $\gamma$-ray emitting
PWN \cite{Aha05a,Aha06b}. Such detections imply one or more of the following properties (see also \cite{dej05a}): 

\begin{itemize}
\item The overall (total) wind magnetization parameter of the PWN $\sigma_{\rm tot}$ may be much less
than unity (i.e. a particle dominated wind as discussed by \cite{Che04}, 
whereas \cite{dej95} considered evolved PWN in equipartition. If the PWN is well below equipartition ($\sigma_{\rm tot}\ll 1$),
synchrotron losses on the VHE emitting particles will be relatively small, leading to an intrinsically bright
VHE source;

\item Rapid expansion of the PWN during its early phases of high power input from the
pulsar (such as G0.9+0.1 and the PWN of PSR\,B1509-58 \cite{Che04})
results in a relatively low field strength in the PWN and hence 
the survival of the majority of VHE emitting electrons 
since early epochs. If the progenitor was part of an OB association,
the combined stellar wind would blow a cavity so that the 
PWN expands nearly uninhibited for tens of kyr. The
magnetic field strength in such an expanded PWN
can in principle drop below a few microgauss, so that 
synchrotron losses become less important relative to IC, ending
up in a ``dark VHE source'', since we can no longer
rely on synchrotron emission to provide a multiwavelength counterpart.
One such possibility is HESS\,J1303-631 which may be associated with
Cen OB as reported by \cite{Aha05b}.

\item The ideal condition (which includes the first two conditions) 
is to have the radiation lifetime $\tau(E_\gamma)$ of VHE radiating particles comparable to, or longer
than, the age $T$ of the system, and therefore surviving even the earlier epochs
when the field was stronger, so that the total amount of energy in electrons in the
PWN is a significant fraction of the maximal rotational kinetic energy of the neutron star 
$I\dot{\Omega}_0^2/2$ at birth. Only adiabatic
losses are then the main source of losses. In this case we do not expect to
see an energy dependence of the PWN size with changing $\gamma$-ray energy --
a well known phenomenon for PWN where the lifetime of particles exceeds the
age of the system. The predicted VHE $\gamma$-ray flux will then depend on the
birth period of the pulsar.  

\end{itemize}

The discovery of VHE $\gamma$-rays from HESS\,J1825-137 \cite{Aha05c} 
and Vela X \cite{Aha06a}, both offset from their pulsars, confirm
predictions of \cite{van01,Blo01,Gae03} 
that anisotropic reverse shocks can offset PWN from their original
positions. We will discuss this futher in this review paper.  
This conclusion led \cite{dej05a} to propose that 
a systematic investigation of middle-aged pulsars at the edges of resolved 
\textit{center-filled} VHE $\gamma$-ray sources, combined with follow-up radio and 
X-ray imaging information, should result in the identification of a new class of 
PWN, with Ground-Based Gamma-Ray Astronomy taking the lead in this new direction. 
{\it GLAST} operations at its highest energies (where the
angular resolution is best) is also expected to make a contribution to this field.
In fact, with {\it GLAST} we expect to see a population of electrons with
ages even longer than those seen by {\it HESS}.

\cite{Lem05} were the first to search for
molecular clouds (based on CO and HI data) associated with such {\it HESS} sources, and in the case
of HESS\,J1825-137, they found a cloud at the same kinematic distance to
its associated pulsar PSR\,B1823-13. This cloud also has the correct orientation relative to
the pulsar and $\gamma$-ray center of gravity to explain the observed offset
in terms of an early reverse shock from the cloud location. A more detailed
study on this topic was performed in the thesis of Lemi\`ere \cite{Lem06} whose
study have revealed this new class of offset, filled center VHE PWN.

\section{Energy scales and lifetimes of X-ray synchrotron and VHE IC emitting electrons \label{sec:2}}
The IC scattering of VHE electons on the 2.7K CMBR radiation is relatively efficient,
since most scattering relevant to the observed VHE $\gamma$-ray range is in the Thomson limit.
Far infrared photons from galactic dust at an average temperature of 25K
tend to preselect lower energy electrons where the Thomson limit still applies, whereas
the cross-section for IC scattering decreases towards higher energies in the
Klein-Nishina limit. The net effect is a steepening in the observed spectrum
if far IR photons due to dust grains dominate, whereas the spectral index for 
IC in the Thomson limit is the same as that for synchrotron radiation by
the same spectrum of electrons in a magnetic field. The reader is
referred to the treatment by \cite{Blu70} on these topics.

In the prediction of a population of VHE $\gamma$-ray emitting PWN due to CMBR and far IR photons
from dust, \cite{dej95} remarked that the effect of dust would become
more dominant towards the galactic center region. The most interesting application of this is the convincing 
prediction that a significant fraction of the galactic center H.E.S.S. source (HESS J1745-290) at the location
of Sgr A* is due to a pulsar wind nebula: The relativistic electrons of this PWN suffers 
severe energy losses as a result of inverse Compton scattering on the dense IR radiation field
\cite{Hin07}.

In the scaling equations below we will focus on the contribution from the CMBR alone,
but refer the reader to \cite{dej95,dej05b} for first order analytical
Klein-Nishina corrections when including the dust component (also based on \cite{Blu70}. 

The required electron energy in a transverse magnetic field of strength $B=10^{-5}B_{-5}$ G,
to radiate synchrotron photons of mean energy $E_{\mathrm keV}$ (in units of keV) is given by 
\begin{equation}
E_e=(70\;{\mathrm TeV})B_{-5}^{-1/2}E_{\mathrm keV}^{1/2}.
\label{EeX}
\end{equation}
Similarly, the mean electron energy required to inverse Compton scatter the CMBR seed photons
to energies $E_{\mathrm TeV}$ (in units of TeV) is typically lower at
\begin{equation}
E_e=(18\;{\mathrm TeV})E_{\mathrm TeV}^{1/2}.
\label{Eeg}
\end{equation}
The mean synchrotron photon energy (in units of keV) can be written in terms of the
mean IC photon energy (following scattering on the CMBR) $E_{\mathrm TeV}$ (in units of TeV)
\begin{equation}
E_{\mathrm keV} = 0.06 B_{-5}E_{\mathrm TeV}.
\end{equation}

If we define the synchrotron lifetime ($\tau=E/\dot{E}$)
of a VHE electron in a transverse magnetic field $B_{\perp}$, scattering cosmic microwave background (CMBR)
photons to energies $E_{\gamma}=10^{12}E_{\rm TeV}$\,eV (in the
Thomson limit) can be shown to be 
\begin{equation}
\tau(E_\gamma)\sim (4.8\;{\rm kyr})\left(\frac{B_{\perp}}{10^{-5}\,{\rm G}}\right)^{-2}E_{\rm TeV}^{-1/2}.
\label{tau:g}
\end{equation}
Note that the constant above depends on the assumed pitch angle distribution of the
electron relative to the magnetic field direction. For an isotropic distribution
this constant becomes 7.6 kyr. Other constants will also be
obtained if we define the final electron energy to be 
an arbitrary fraction of the initial energy. 
Assuming Eqn (\ref{tau:g}), the corresponding lifetime of keV emitting electrons would be shorter:
\begin{equation}
\tau(E_{\mathrm{X}})=(1.2\;{\rm kyr})\left(\frac{B_{\perp}}{10^{-5}\,{\rm G}}\right)^{-3/2}E_{\rm keV}^{-1/2},
\label{tau:X}
\end{equation}
where $E_{\rm keV}$ is again the synchrotron photon energy in units of keV.
Once again, the constant 1.2 kyr increases to 1.6 kyr if we assume isotropic pitch angles.
In very extended nebular MHD flows it is possible to get field strengths
below the typical $3\mu$G ISM value, in which case we should include
inverse Compton energy losses on the 2.7K CMBR. Following \cite{Aha06c}
we also include a Klein-Nishina suppression factor (relative to the Thomson limit) 
of $2/3$ for the H.E.S.S. range to give a lifetime for TeV emitting electrons of 
\begin{equation}
\tau(E_\gamma)\sim \frac{(100\,{\rm kyr})}{[1+0.144(B_{\mu{\rm G}})^2]E_{\rm TeV}^{1/2}},
\label{tauglow}
\end{equation}
where $B_{\mu{\rm G}}$ is now the field strength in unit of microgauss. This hints
at a terminating VHE lifetime of 100 kyr.
Note however that the X-ray synchrotron surface brightness in such extended low-field
environments may be well below detectable levels, in which case a PWN will only be visible
in the $\gamma$-ray domain (via the IC process), whereas no radio, optical and/or X-ray (synchrotron)
plerionic counterpart is found. {\bf Furthermore, the parent neutron star's 
thermal and/or non-thermal X-ray component(s) may also no longer be visible, since the lifetime of VHE emitting electrons
may even exceed the neutron star's cooling and non-thermal radiating timescales!}  

Such $\gamma$-ray detections are expected to contribute to the growing population of 
{\it Unidentified Gamma-Ray Sources} or {\it Dark Accelerators/Sources}, but given
our growing knowledge through experimentation and theory, we should eventually be 
able to lift these PWN to the status of ``{\it gamma-ray pulsar wind nebulae} without multiwavelength counterpart.'' 

\section{Particle acceleration at PWN shocks\label{sec:3}}

Although electrons and positrons from magnetospheric electromagnetic cascades 
escape with relatively low energy from the
pulsar light cylinder, where the ratio $\sigma_{L}$ of electromagnetic energy density
to particle energy density must be much larger than unity ($\sigma_L \sim 10^4$ for the Crab pulsar \cite{Cor90}
this $\sigma$ parameter must reduce drastically beyond the light cylinder for two reasons:
(i) \cite{Wil78} have shown that the Crab pulsar's radio pulses would have been
smeared unless the Lorentz factor of the $e^{\pm}$ wind exceeds $10^4$; (ii)
Observations at the pulsar wind shock indicate that this ratio must have reduced to
$\sigma \sim 3\times 10^{-3}$ for the Crab \cite{Ken84,Bog05},
but in the range 0.01 to 0.1 for the Vela PWN shock \cite{Sef03,Bog05}. 
The reader is referred to \cite{Aro04} for a review of this particle energization process. 

The first observational evidence of a pulsar wind nebular shock was seen in the Crab Nebula,
where the observation of the underluminous region within $\sim 10$ arcsec from the pulsar
made it natural for \cite{Ree74} to associate the pulsar wind shock ($R_s\sim 0.1$ pc) with the
boundary of this region. 

Even though energy is converted from fields to the bulk of the particle population between $R_L$ and $R_s$, 
we require an additional mechanism to accelerate electrons and positrons to ultrarelativistic energies with an electron
spectral index of $\sim 2$ (see Section \ref{sec:4} for observational evidence). Recently \cite{Ama06}
have shown that the fractional energy content in the accelerated $e^{-}-e^{+}$ component increases with increasing
energy content in upstream (unshocked) ultra relativistic ions. These authors also found that the ratio of
accelerated $e^{+}$ to $e^{-}$ components become significantly more than unity as the upstream wind energy flux
becomes ion dominated. The reader is also referred to \cite{Ama06} for a review of the literature on ion mediated 
lepton acceleration in pulsar wind shocks.

Another fundamental question regarding pulsar wind nebulae is the maximum
energy to which $e^{-}-e^{+}$ can be accelerated at a pulsar wind shock to produce the 
observed high energy synchrotron and inverse Compton radiation downstream of the shock?
We will attempt to answer this question without restricting ourselves to the microphysics of the
acceleration process.

This limit for low field strength pulsar wind shocks was first discussed by \cite{Har90} who considered
the gyroradius limit and by \cite{dej96a} for synchrotron limited acceleration as discussed below. 

\subsection{The synchrotron limit\label{sec:syn}}
For any ion gyro resonant (as discussed before) or Fermi I type acceleration process in a relativistic pulsar wind shock, 
electrons and/or positrons can be accelerated at a rate as fast as the gyroperiod, as reviewed by \cite{dej96a}. The latter authors then balanced this rate with synchroton losses for relatively strong fields to derive a maximum characteristic (cutoff) synchrotron energy 
(in the lab frame) of 
\begin{equation}
E_{\gamma}({\rm max})=\left(\frac{3}{4\pi}\right)^2 \left(\frac{hc}{r_0}\right)=(25\,{\rm MeV}) 
\label{e_crab}
\end{equation}
for electrons/positrons in such relativistic shocks.
This limit reproduces the observed characteristic synchrotron cutoff energy of $\sim 25$ MeV for the 
Crab Nebula as seen by {\it COMPTEL} and {\it EGRET} \cite{dej96a}. 
The corresponding 90\% confidence interval for this cutoff is 15 MeV to 70 MeV.

We note that \cite{Ach01} approached this problem from a totally different perspective
by calculating the electron cycle times upstream and downstream of a relativistic shock.
They finally arrived at exactly the same expression (Eqn \ref{e_crab}), or Eqn. (A2) of \cite{dej96a}
within a factor of $2\pi$. \cite{Aha00} also extended this limit to account for synchrotron limited
acceleration of protons and electrons by considering an acceleration timescale given by
a scale factor $\eta$ times the gyroradius relative to $c$, so that Eqn (A2) of \cite{dej96a}
is retrieved if we set $\eta=2\pi$ (gyroperiod timescale). His maximum for electrons is then 
\begin{equation}
E_{\gamma}({\rm max})=(160\,{\rm MeV})\eta^{-1}\, 
\label{e_eta}
\end{equation}
similar to \cite{Ach01}. EGRET observations of the Crab Nebula are however consistent with
Eqn \ref{e_crab}, or, $\eta\sim 2\pi$ \cite{dej96a}.

\subsection{The gyroradius limit}

For lower field strengths (i.e. PWN evolved beyond the Crab-like phase), where synchrotron losses no longer
constrain the maximum electron energy, we generalise the gyroradius limit of \cite{Har90},
which states that the highest energy particles must still be contained in the shock at $R_s$ while
participating in the acceleration process. In other words,
the gyroradius of particles with energy $E_{\rm max}$ should be significantly smaller than $R_s$, or,
\begin{equation}
 r_L =\frac{E_{\rm max}}{eB_s} =\epsilon R_s < R_s,
\end{equation}
where $\epsilon<1$ is the required fractional size of $R_s$ for containment during 
any Fermi or gyro resonant type of acceleration to provide the 
bright synchrotron radiation downstream of the pulsar wind shock.
This result does not depend explicitly on the acceleration timescale, but
relatively fast acceleration (relative to escape and radiation losses) will result in
maximal values of $\epsilon <1$, until containment becomes a problem.

We can again generalise this expression by adding 
the charge number $Z$ ($=1$ for electrons and positrons) for the possible 
acceleration of ions to give
\begin{equation}
E_{\rm max} = Ze\epsilon B_sR_s.
\label{emax_1}
\end{equation}

Eqn (\ref{emax_1}) converges to Equation (52) of \cite{Ach01}
if we set the abovementioned containment factor $\epsilon=(\ell_c/3r_L)^{1/2}<1$, where $\ell_c$ is the coherence
length of the field, if the latter is assumed to be randomly oriented. We, however, do not
specify the exact physics of acceleration so that $\epsilon$ is kept as a free parameter, which is constrained
to be less than unity from a general containment principle.
 
We apply this constraint to PWN by using the \cite{Ken84} formalism to write
the post-shocked field strength in terms of the pulsar spindown power and parameters related to the
pulsar wind shock to give 
\begin{equation}
B_s  =  \kappa \left[  \frac{\sigma \dot{E}}{(1+\sigma)c} \right]^{\frac{1}{2}}\frac{1}{R_s},
\end{equation} 
with $\sigma$ the wind magnetization parameter ($\sigma \equiv$ electromagnetic energy density
to particle energy density at $R_s$) and $\dot{E}$ the spindown power of the pulsar.
The magnetic compression ratio $1<\kappa<3$ \cite{Ken84} depends on the strength
of this relativistic shock (and hence $\sigma$). For strong shocks ($\sigma \ll 1$) we have
$\kappa\sim 3$, whereas $\kappa\sim 1$ for weaker Vela-like shocks \cite{Ken84}, 
where $\sigma\sim 0.01$ to 0.1 \cite{Sef03,Bog05}.

The expression for the maximum particle energy for remnants with field strengths
weaker than the Crab (i.e. where radiation losses do not limit the energy)
can now be written in terms of $\sigma$ and $\dot{E}$,
without having to know where the shock is located (since $R_s$ cancels):
\begin{equation}
E_{\rm max} = \epsilon \kappa e \left(\frac{\sigma}{1+\sigma}\frac{\dot{E}}{c}\right)^{1/2} = 
(110\,{\rm TeV})\kappa \left(\frac{\epsilon}{0.2}\right)\left(\frac{\sigma}{0.1}\dot{E}_{36}\right)^{1/2},
\label{emax}
\end{equation}
where the spindown power has been rescaled in terms of a typical Vela-like value of $\dot{E}=10^{36}\dot{E}_{36}$ ergs/s
and a Vela-like $\sigma\sim 0.1$. However, according to
\cite{Har90} the maximum energy is equal to the polar cap potential, in which
case the term $\epsilon\kappa\sigma/(1+\sigma)$ should drop away (i.e. no field compression
and magnetisation). In this case the maximum energy for a Vela-like pulsar would be $\sim 350$ TeV
if $\epsilon=0.2$. 

The discussion of this maximum is very relevant from an observational viewpoint, since
the highest photon energy at any location in the PWN will always be bounded by such a quantity, which should evolve with time
as a result of pulsar spindown. 

\subsection{PSR\,B1929+10: a challenge for particle acceleration in PWN shocks.}
The aforemetioned maximum electron energy imposes an important boundary condition when 
modelling the extended X-ray and $\gamma$-ray emission
from PWN, since this maximum will suffer synchrotron and adiabatic losses during electron transport
to the outer edge of the PWN. While this long-term transport process is taking place, the spindown power 
will also be decreasing with time, resulting in a decrease
of $E_{\rm max}$ at $R_s$ for freshly injected electrons (making compact nebular X-rays) as well.
For the sake of simplicity we will assume that the maximum electron energy is set
by the pulsar polar cap potential \cite{Har90}.

The detection of an X-ray synchrotron trail from PSR\,B1929+10 by \cite{Wan93,Bec06}
from X-ray observations should challenge the
maximum electron energy set by Equation (\ref{emax}), since the spindown power $\dot{E}=3.9\times 10^{33}$ ergs/s
is low compared to even Vela-like pulsars. \cite{Bec06} considered
the case where the time $t=R_t/V_p$ (with $R_t$ the length of the trail seen in X-rays
and $V_p$ the pulsar proper motion velocity) is equal to the synchrotron lifetime $\tau_s$,
which can be several thousand years for a trail field strength $B<10\mu$G.
Even if we assume that the electrons never lost any energy along their transport
from $R_s$ to the present weak-field trail location, we find from Equation (\ref{emax}) that the maximum
electron energy is given by
\begin{equation}
E_{\rm max}({\rm B1929})\sim (20\,{\rm TeV}) \left(\frac{\epsilon}{0.2}\right),
\label{emax_1929}
\end{equation}
whereas the corresponding characteristic synchrotron energy in a nebular magnetic field, 
normalised to a strength of $B_{-5}=B/10\mu{\rm G}$, would be given by
\begin{equation}
h\nu_{\rm max}({\rm 1929})< (300\,{\rm eV})B_{-5}\left(\frac{\epsilon}{0.2}\right)^2.
\label{hv_1929}
\end{equation}
This limit is clearly in the soft X-ray band, falling short to explain synchrotron emission
up to 10 keV as observed.

If the pulsar has spun down significantly during a time $t$ into the past,
then we are looking at relic electrons in the trail,
but accelerated at the pulsar wind shock when the age of the pulsar was equal to
$T-t$, where $T$ is the current true age. Assuming a pulsar braking index of $n=3$,
the {\it retarded spindown power} would then be 
\begin{equation}
\dot{E}=\frac{\dot{E}_0}{(1+(T-t)/\tau_0)^2},
\end{equation}
where $\dot{E}_0$ is the spindown power at birth and $\tau_0=P_0/2\dot{P}_0$ the
characteristic age at birth when the initial spin period was $P_0$ and period
derivative $\dot{P}_0$. For those not familiar with the concept of a {\it retarded spindown power}:
The author borrowed this concept from electrodynamics where the term {\it retarded potential}
is used to describe radiating systems, where the changing vector potential seen by the observer is the result of particle
movement some time in the past. Also, for those not familiar with the concept of a {\it braking index}:
The spindown power of a pulsar can be written as a function of the rotational frequency $\Omega$ as
$-\dot{E}=K\Omega^{n+1}$, where $K$ depends on several neutron star properties and $n$ the braking index.
For magnetic dipole radiation and energy losses via particle outflow through the pulsar polar cap
as in the well-known Goldreich-Julian model, $n=3$.
However, even this retarted spindown may not solve the
problem of PSR\,B1929+10 since the age $T=3\times 10^6$ years (i.e. already too large), so that
quantity $(T-t)/T\sim 1$ does not give us any advantage.

We can therefore only speculate about possible explanations for the
existence of X-ray synchrotron photons from the trail of PSR\,B1929+10: One possibility
is re-acceleration due to adiabatic compression in the bow shock. A detailed
discussion of this is however beyond the scope of this general review paper, except to mention that
more X-ray observations,
as well as future ground based VHE $\gamma$-ray observations are important
to characterise this important laboratory for particle acceleration. 
Whereas X-rays measure the convolution of the electrons
with the field strength in the trail, the VHE $\gamma$-rays would
directly probe the particle population via the IC scattering of this relic component
on the CMBR and known far infrared photons from galactic dust grains.
 
\section{The energy dependent cooling radius of a PWN \label{sec:4}}

\begin{figure}
\centerline{\psfig{file=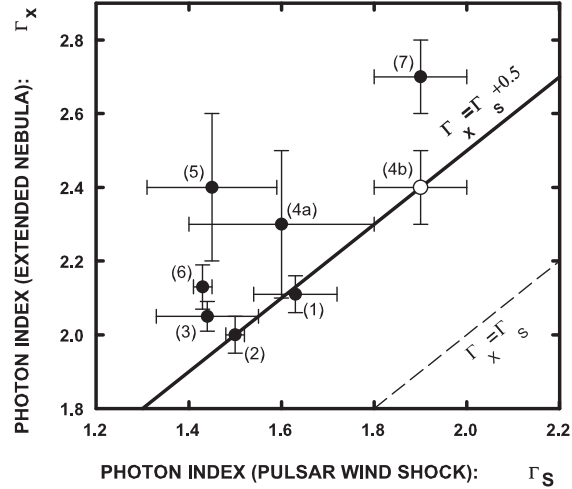,width=8cm} }
\caption{Plot of the X-ray spectral index $\Gamma_X$ around $\sim 1$ keV at the maximal 
observable radius (or ``extended nebula'') versus the spectral index $\Gamma_s$ for the same
energy range near the pulsar wind shock. The dashed line indicates $\Gamma_X=\Gamma_s$
whereas the solid line represents the line $\Gamma_X=\Gamma_s+0.5$ 
expected for $dE/dt\propto -E^2$ (e.g. synchrotron) cooling if we assume that the
maximum observable photon energy is still above the upper spectral edge of the observations
in the extended nebula (typically $>10$ keV). References: (1) Crab \cite{Got02},
(2) Vela \cite{Man05,Mar97}, 
(3) PSR B1509-58 \cite{Gae02}, (4a) G18.0-0.9 \cite{Gae03},
(4b) HESS\,J1825-137 \cite{Aha06c},
(5) G0.9+0.1 \cite{Por03}, (6) G21.5--0.9 \cite{Mat05}, \& 
(7) 3C58 \cite{Boc01,Sla04}
}
\label{fig:1}        
\end{figure}

In Figure \ref{fig:1} we summarise the observed spectral steepening with radius $r>R_s$, where
$R_s=d\theta_s$ is the pulsar wind shock radius, $d$ is the distance between earth and the pulsar, 
and $\theta_s$ is the angular distance between the pulsar and its wind shock as seen on the sky:  
If $\Gamma_s$ is the photon spectral index (as seen in X-rays) at the pulsar wind shock, a steepening
$\Gamma_x>\Gamma_s$ corresponding to $r>R_s$ is observed as a result of radiative losses. This
effect is mostly seen in the X-ray (synchrotron) domain, but may be less so in the VHE (IC) domain
where lower energy electrons (producing the VHE $\gamma$-rays) suffer less radiative losses compared
to the synchrotron emitting electrons. 
  
Theoretically we would expect a convergent value of $\Gamma_X=\Gamma_s+0.5$ as a result
of radiative (mostly synchrotron) cooling, since the energy loss rate scales as electron energy squared.
It is clear that the points all lie either on this (solid) line in Figure \ref{fig:1},
or above it. The reason for the latter is because the spectral cutoff
associated with the highest energy electrons at the observed angular radius $\theta_X$ 
may only contribute to energy bins below the upper spectral edge used in the
analysis, in which case the spectral index $\Gamma_X$ may asymptotically diverge
to relatively large numbers near $\theta_X=r/d$. However, 
even this information is useful in sophisticated models
where the maximum electron energy in the PWN shock serves as one of the inputs.
This can help us to constrain the maximum electron energy 
at $\theta_s$ in time dependent models which takes the full evolution into account.

Any PWN has a terminating radius $\theta_{\rm PWN}$, which pushes against the SNR ejecta, swept-up
gas or ISM. This edge is usually seen in radio (if the PWN is also detectable in radio), 
because the lifetime of radio synchrotron emitting electrons
is longest and they survive in the oldest expanding volume at the radial distance $\theta_{\rm PWN}$. 
Suppose we define the angular radius $\theta_{1/2}$, with 
$\theta_s< \theta_{1/2}\leq \theta_{\rm PWN}$, as that radius where $\Gamma_X=\Gamma_s+0.5$
is reached. 
We then define the energy dependent scaled cooling radius as
\begin{equation}
\xi_{1/2}\equiv \frac{\theta_{1/2}}{\theta_{\rm PWN}},\;\;{\rm where}\;\;\frac{d\xi_{1/2}}{dE_e}<0.
\label{xi}
\end{equation}
The scaled radius $\xi_{1/2}$ should thus decrease towards increasing electron energies $E_e$.
Although this is well-known, there is little experimental data to support this:
Detectors with good angular resolution to resolve PWN also have limited
bandwidth (typical 0.5 to 10 keV), but if the statistics are good enough,
observers should consider splitting the data into two energy bands of equal
statistics to identify a possible shift in $\xi_{1/2}$ between the two
energy bands. 

In the Section \ref{hess1825} we will review the H.E.S.S. detection
of the PWN G18.0-0.7 = HESS J1825-137 associated with the
Vela-like pulsar PSR B1823-13, which clearly shows a similar
steepening of the photon index with radius. 

\section{{\it Pleres pera} or ``filled bags''\label{vb}}
The most fundamental principle in PWN flows is that the pulsar wind slows
down from relativistic to relatively low expansion velocities as a result of the confining pressure.
This slow down typically occurs after the pulsar wind has been shocked at
some distance $R_s$ from the pulsar (see e.g. \cite{Ken84}. 
The decelerating post shock ($r>R_s$) pulsar wind flow velocity ${\bf V}$ would typically be radial, i.e.
${\bf V}\sim V_r{\bf e_r}$. Furthermore, the associated magnetic field at $r>R_s$ would be described by
the equation (see e.g. \cite{Ken84}
\begin{equation}
\nabla\times ({\bf V}\times {\bf B}) \sim 0,
\label{toroidal}
\end{equation}
in which case ${\bf B}\sim B{\bf e_\perp}$: The field direction would also be approximately perpendicular 
relative to the radial flow direction. 
The reader is probably familiar with the {\it Chandra} image of the Crab Nebula and other X-ray plerions,
showing exactly the toroidal (azimuthal) structures implied by this vector Equation.  

Given this introduction, we now raise two questions: 
(i) Is it possible for a high energy particle to overtake this convective flow as a result of
diffusion? (ii) Under which conditions would it be possible for such a particle to
escape through the boundary of a PWN? We will therefore compare radial diffusion
versus radial convective flow, where the diffusion is {\em perpendicular} to the 
(e.g. toroidal) magnetic field line. 

The most general form of the perpendicular diffusion coefficient is given by
\begin{equation}
\kappa_{\perp}=\frac{1}{3}\lambda_{\perp}c,
\end{equation}
where $\lambda_{\perp}$ is the mean free path for scattering in a direction which is 
perpendicular to the field line direction (i.e. cross field diffusion). The case we consider here 
therefore corresponds to diffusion in the radial direction relative to the pulsar. We
parameterize this quantity further by writing it as a factor $f$ times the particle gyroradius
$\rho_L$, so that 
\begin{equation}
\lambda_{\perp}=f(\Omega \tau)\rho_L,
\label{f}
\end{equation}
where $\Omega$ is the particle gyrofrequency and $\tau$ is the mean time between scatterings.
The {\em Bohm limit} corresponds to $\Omega \tau\sim 1$ and $f(\Omega\tau)= 1$. Assuming hard sphere
scattering, it can be easily shown for both weak scattering ($\Omega\tau \ll 1$) and 
strong scattering ($\Omega\tau \gg 1$) that $f(\Omega\tau)\ll 1$ \cite{Ste88}.
Thus, under the assumption of hard sphere scattering, the mean free path against diffusion perpendicular
to a field line is always less or equal to the particle gyroradius. This is intuitively expected:
It is difficult for charged particles to cross field lines - a principle we have learned from undergraduate days!
 
Scaling a PWN to a distance of $d=1d_{\rm kpc}$ kpc with an age of $\tau=20$ kyr and an
associated field strength of $B=3\mu$G, we arrive at an angular spread due to diffusion
of VHE electrons scattering CMBR photons in the Thomson limit to VHE $\gamma$-rays of energy
$E_{\rm TeV}$ (from Eqn \ref{Eeg})
\begin{equation}
\theta_{\rm diff}(e^{\pm})=0.07^{\circ}d_{\rm kpc}^{-1}
\left[(\frac{f(\Omega \tau)}{0.1})(\frac{3\,\mu{\rm G}}{B})(\frac{\tau}{10\,{\rm kyr}})\right]^{1/2}E_{\rm TeV}^{1/4}.
\label{theta_diff}
\end{equation}

Thus, the toroidal field line structures in PWN tend to contain relativistic particles much
better than would have been the case if the field line structures had radial components, in which
case the parallel diffusion coefficient is relatively large ($\lambda_{||}\gg \rho_L$). 
In the latter case
much larger cosmic ray type diffusion coefficients would have been appropriate as employed by \cite{Aha97}
for PSR\,B1706--44, so that relic charged particles accumulated over the source lifetime  
would have escaped much more easily from the convective plasma, with the latter reflected by the radio morphology.

We therefore conclude that PWN act as well-contained ``filled bags'' (or {\it pleres pera} in Ancient Greek) 
with high integrity against diffusion losses over Vela-like lifetimes. Such sources will expand convectively
until the PWN pressure becomes small enough so that particles start to leak into the interstellar medium.
The same principle is also expected to hold for ultrarelativistic ions injected into the PWN over the lifetime of the PWN:
If $E_i$ is the total energy per ultra relativistic ion with charge $X$, the spatial ion spread due to
diffusion alone is then
\begin{equation}
\theta_{\rm diff}({\rm ion})=0.05^{\circ}d_{\rm kpc}^{-1}
\left[(\frac{1}{Z})(\frac{f(\Omega \tau)}{0.1})(\frac{3\,\mu{\rm G}}{B})(\frac{\tau}{10\,{\rm kyr}})(\frac{E_i}{10\,{\rm TeV}})\right]^{1/2}.
\label{theta_ion}
\end{equation}

If we claim that several of the unidentified filled-center H.E.S.S. sources near Vela-like pulsars
in the galactic plane are PWN, diffusion would spread them by undetectable amounts 
(given the {\it HESS} angular resolution of $\sim 0.07^{\circ}$) relative to convective
sizes $\theta_{\rm PWN}$. For example, scaling the $\sim 2$ degree convective size of the Vela X PWN to a distance
of 1 kpc gives $\theta_{\rm PWN}=0.3^{\circ}d_{\rm kpc}^{-1}$, which is large compared to the diffusive size.

\section{HESS J1825-137 and the Three Princes of Serendip \label{hess1825}}

The serendipitous discovery of the source HESS J1825-137 as part of the Galactic plane
H.E.S.S. survey \cite{Aha05a} serves as a good example 
of the correct use of the word {\it serendipity} as coined by Horace Walpole in the 18th Century, based
on the old Persian tale of the {\it Three of Princes of Serendip}. In this story the rewards did
not come at the time of discovery, but only later.
We will also identify three main discoveries following the
collection of sufficient statistics on this source.

The first H.E.S.S. observations of this region occurred as part of a
systematic survey of the inner Galaxy from May to July 2004 (with 4.2
hours of exposure within $2^{\circ}$ of HESS\,J1825--137).  Evidence
for a VHE $\gamma$-ray signal in these data triggered re-observations
from August to September 2004 (5.1 hours), resulting in a significance
of $13\sigma$. This led to the announcement by the H.E.S.S. Collaboration
\cite{Aha05c} of a possible association 
of this source with the Vela-like pulsar PSR B1823-13 and its associated
PWN G18.0-0.7 as identified in X-rays by \cite{Gae03}. 

\begin{figure}
\centerline{\psfig{file=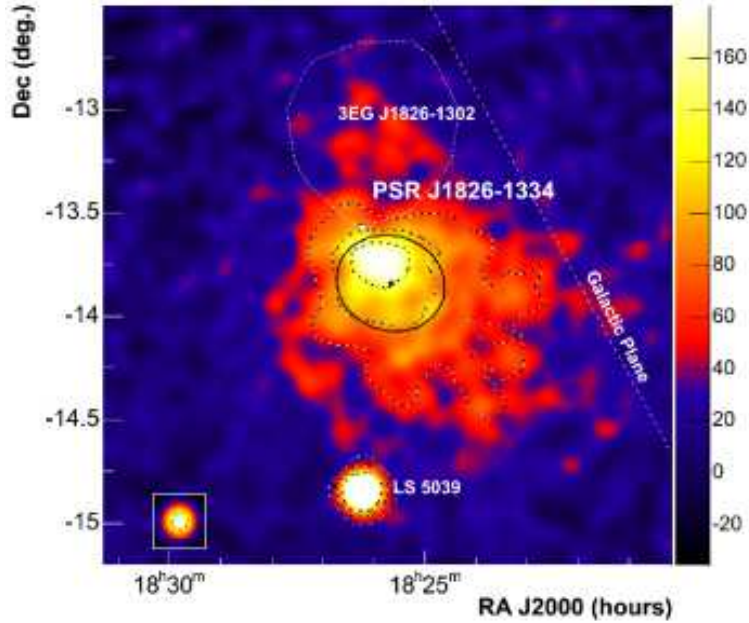,width=10cm,clip=} }
\caption{Acceptance-corrected smoothed map of HESS J1825-137 \cite{Aha06c}
showing the extended emission relative to PSR B1823--13=PSR J1826--1334. 
The dotted line indicates the 95\% confidence contour of the unidentified
EGRET source 3EG J1826--1302. The point source at the bottom corresponds
to the microquasar LS 5039. (Figure from \cite{Aha06c}.)
}
\label{fig:2}        
\end{figure}

Further observations during 2005 resulted in improved statistics 
to study the energy dependent morphology \cite{Aha06c}.
One of the main reasons for this was to get full orbital coverage
on the source LS 5039 as seen in Figure \ref{fig:2}. The total lifetime
then increased to 52.1 hours with a significance of $34\sigma$.

The three main discoveries with respect to HESS J1825-137 are the following:

\subsection{The anomalously large size of HESS J1825-137 and its implied SNR shell}

Since the X-rays already show the effect of full cooling from a
photon index of $\sim 1.6$ to $\sim 2.3$ (Figure \ref{fig:1})
within a distance of $\sim 5$ arcmin from the pulsar \cite{Gae03}, 
we would expect that the X-ray PWN G18.0--0.7 already reached it
terminal size.  If this is not the terminal size, further cooling
well beyond 5 arcmin should then result in $\Gamma_{\rm X}\gg 2.3$
as discussed in Section \ref{sec:3} accompanied by the loss of X-ray statistics.

The VHE $\gamma$-ray size is however $\sim 1^{\circ}$ as seen from Figure \ref{fig:1},
which is much larger than G18.0--0.7 and the anomalously large size of the pulsar wind nebula can be explained
if the pulsar PSR\,B1823--13 was born with a relatively large initial spindown power and
braking index $n\sim 2$, provided that the SNR expanded into the hot ISM
with relatively low density ($\sim 0.003$ cm$^{-3}$). This pulsar is a 101 ms evolved
pulsar with a spin-down age of $T=2.1 \times 10^{4}$ years
for an assumed braking index of $n=3$ \cite{Cli92} and in these properties very similar to the Vela
pulsar. It is located at a distance of $d=3.9\pm 0.4$ kpc \cite{Cor02}.

The average VHE $\gamma$-ray radius of the PWN of $\sim 0.5^{\circ}$ corresponds
to $R_{\gamma}=35d_4$ pc for a distance of $d=4d_4$ kpc. Since the observed ratio of
SNR forward shock radii to PWN radii are all a factor 4 or larger \cite{van01},
the expected SNR forward shock radius $R_{\rm SNR}> 140d_4$ pc, making this
one of the largest expected SNR in our galaxy. de Jager et al. \cite{dej05c} derived a constraint
on this forward shock radius, which can be stretched to a value of

\begin{equation}
R_{\mathrm SNR}=(120\,{\mathrm pc})~
\left(\frac{E_{\mathrm SN}}{3\times 10^{51}{\mathrm ergs}}\frac{0.001\,{\mathrm cm^{-3}}}{N}\right)^{0.2}
\left(\frac{1}{n-1}\right)^{0.4}.
\label{r_snr}
\end{equation}


Finally, returning to the question about the apparent discrepancy between
the VHE and X-ray sizes: We can achieve the observed ratio of 1 degree
(VHE) relative to 5 arcmin (X-rays) easily in a low-B environment
(with $E_{\rm TeV}\sim 0.3$ and $E_{\rm keV}=1$) if we 
adopt the equation for conservation of magnetic flux in spherical coordinates
(assuming the steady state solution, giving $RVB$=const), 
and that the radius $R$ is equal to
the expansion velocity $V$ times radiation lifetime. This would give
a field strength in the outer VHE nebula, which is about 3 times
smaller than the field strength in the smaller X-ray nebula. 
In fact, we do need a small field strength in the outer nebula
to allow VHE emitting electrons to survive while producing a
VHE spectral break as observed. The latter will also be covered in this review.

\subsection{The offset PWN in X-rays and VHE $\gamma$-rays}
At the time of the X-ray discovery of G18.0-0.7, it was found that
whereas the uncooled X-ray compact nebula is symmetric around the pulsar, the
extended cooled nebula is offset to the south.
To explain this offset, \cite{Gae03} introduced the reverse shock 
explanation of \cite{Blo01} for Vela X, where hydrodynamical
simulations have shown that SNR expansion into an inhomogeneous medium
would result in a reverse shock returning first from the
region of higher density. After crushing the PWN, the latter is
offset from its original position, resulting in a new center of
gravity. We also observe that the entire VHE image is shifted
relative to the pulsar (Figure \ref{fig:2}) and the same explanation 
for HESS\,J1825-137 was also offered by us in \cite{Aha05c}.

By extending a line from this shifted VHE center of gravity through
the pulsar, \cite{Lem06} discovered a molecular cloud
in CO at a distance from earth, which is consistent with the dispersion based distance
to the pulsar. 
This means that the SNR forward shock most likely struck this cloud,
resulting in a reverse shock returning first to the expanding PWN of PSR B1823-13, 
which then resulted in a predictable offset direction for the VHE center of
gravity. 

Since this process takes of the order of 3 to 10 kyr to offset a PWN,
we can expect several Vela-like PWN (with ages older than 10 kyr) 
to be offset in VHE $\gamma$-rays, since SNR expansion is always 
expected to take place in an inhomogeneous medium (i.e. the
ISM is rarely expected to be homogeneous). 

\subsection{Energy dependent morphology and the cooling break}
\subsubsection{(a) Spectral steepening away from the pulsar}
\begin{figure}
\centerline{\psfig{file=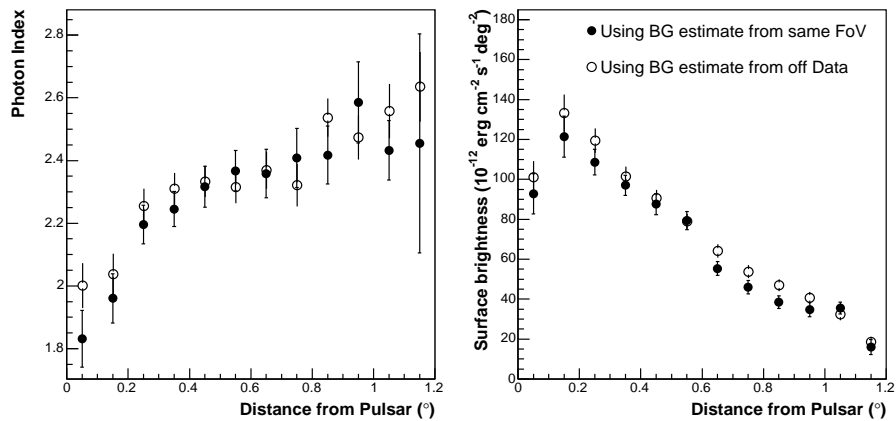,width=12cm,clip=} }
\caption{{\bf Left}: Photon index vs. radius along the sector of brightest emission
from the pulsar as defined by \cite{Aha06c}. {\bf Right}: Relative surface
brightness (all energies) corresponding to the left panel.
}
\label{fig:3}        
\end{figure}

The most astonishing discovery of this source (given the extended statistics)
was the discovery of a steepening spectrum as a function of increasing distance
from the pulsar as described in \cite{Aha06c} as shown on the left
side of Figure \ref{fig:3}, where we see the photon index in intervals of
0.1 degrees (along the sector of brightest emission) increasing from $1.9\pm 0.1$
to $2.4\pm 0.1$ at the outer part. The corresponding surface brightness for the
same slice/sector is shown in the right-hand panel of Figure \ref{fig:3}
and note the peak at a distance of $0.15^{\circ}$ from the pulsar. We will
revisit this feature below.

\begin{figure}
\centerline{\psfig{file=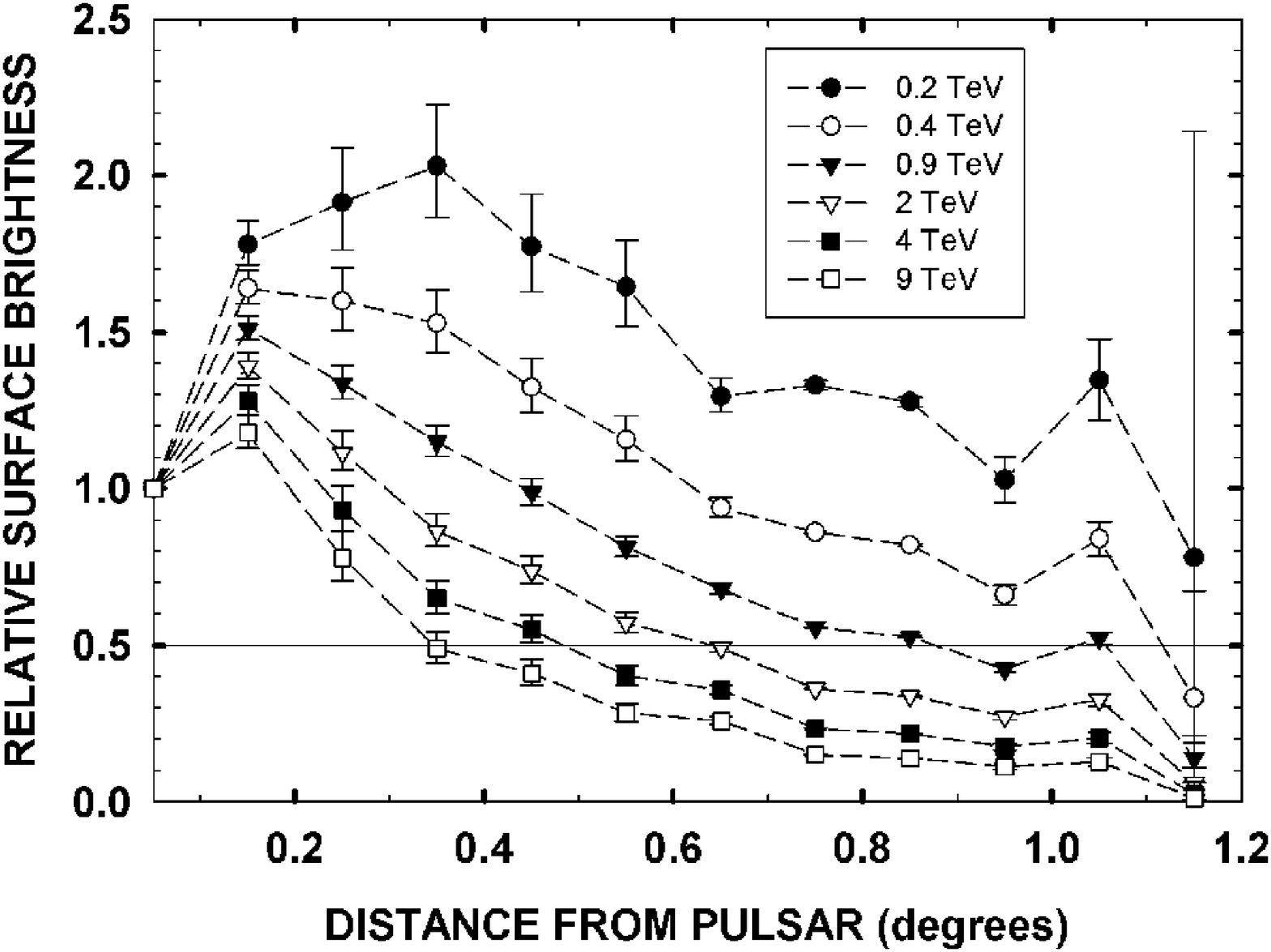,width=10cm,clip=} }
\caption{Relative surface brightness versus distance from the pulsar for energies
between 0.2 TeV and 9 TeV as indicated in the legend. The relative brightness
along the energy scale is correlated (because these points and their errors were derived
from power law fits to individual radial slices), whereas it is uncorrelated along the radial
scale.}
\label{fig:4}        
\end{figure}

We can clearly see the energy dependence of the surface brightness with radius
if we take the power law fits corresponding to each radial interval (from \cite{Aha06c}) 
and calculate the relative surface brightness for energies chosen
within the energy limits of the power law fits. The results are shown in Figure \ref{fig:4}.
Note that these relative surface brightness plots therefore represent smoothed averages
over energy, whereas the radial scale remains uncorrelated. 

In this plot we can see that the size of the source shrinks with increasing energy
and that the peak surface brightness (corresponding to the shifted PWN due to
the effect of the reverse shock) in Figure \ref{fig:3} is located at a radius 
of $\theta_{\rm peak}\sim 0.2^{\circ}$ from the pulsar for $E_{\gamma}\sim 0.2$ TeV 
in Figure (\ref{fig:4}). This peak however 
shifts towards the pulsar for increasing energies (i.e. $\theta_{\rm peak}\leq 0.1^{\circ}$ for 
$E_{\gamma}> 0.9$ TeV). Also, the relative surface brightness drops below the 
50\% level only for $E>0.9$ TeV. 

\begin{figure}
\centerline{\psfig{file=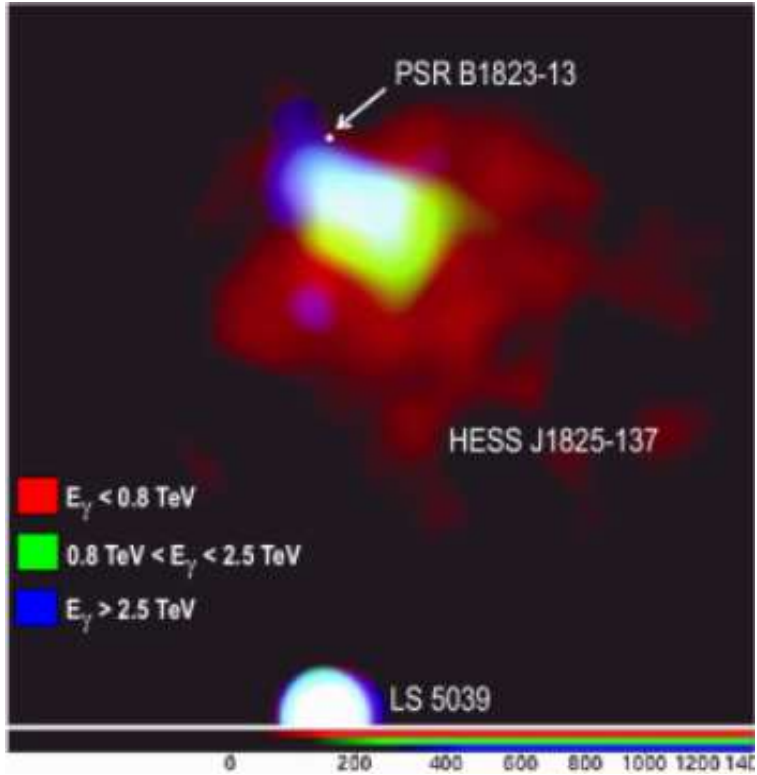,width=10cm,clip=} }
\caption{First color image in the history of Gamma-Ray Astronomy 
showing the energy dependent morphology resolved in the three basic colors
as indicated in the legend. The white point source at the bottom is the $\mu$-quasar LS 5039.
Produced by S. Funk and O.C. de Jager for the H.E.S.S. Collaboration.}
\label{fig:5}        
\end{figure}

With this clear evidence of an energy-dependent morphology we also show
the first color image in the history of Gamma-Ray Astronomy (Figure \ref{fig:5}): 
Defining the three basic
colors RGB as ${\rm R}\equiv [E_{\gamma}<0.8{\rm TeV}]$, ${\rm G}\equiv [0.8{\rm TeV}<E_{\gamma}<2.5{\rm TeV}$
and ${\rm B}\equiv [E_{\gamma}>2.5{\rm TeV}]$, we could combine these colors in a single colour image
showing the extended red image, which shrinks with increasing energy towards the blue nebula
above 2.5 TeV close to the pulsar. Note that the point source LS 5039 shows up as a white
image because of its broad band nature and the fact that it is a point source.

We are now also in a position to compare the scaled cooling radii $\xi_{1/2}$
(Equation \ref{xi}) between X-rays and VHE $\gamma$-rays: 
For X-rays $\xi_{1/2}\sim 5'/1^{\circ}\sim 0.1$,
whereas $\xi_{1/2}\sim 1$ as measured by H.E.S.S. Thus, clearly  
$d\xi_{1/2}/dE_e<0$ as required by Equation (\ref{xi}), where
$E_e$ is the electron energy, which is higher for X-rays than
for VHE $\gamma$-rays, as required by Equations (\ref{Eeg}) and (\ref{EeX})
for a relatively low magnetic field strength as motivated above. 
Furthermore, since $\xi_{1/2}$ is already close to unity for the VHE (IC) domain,
we expect that $\xi_{1/2}$ should become undefined for the GLAST (IC) domain
since the electron lifetime will become longer than the age of the
system for all positions in the PWN -- similar to the radio emission in
the Crab Nebula below the spectral break of $10^{13}$ Hz, which
does not show any cooling effects anymore.

\subsubsection{(b) The cooling break in the total spectrum}
\begin{figure}
\centerline{\psfig{file=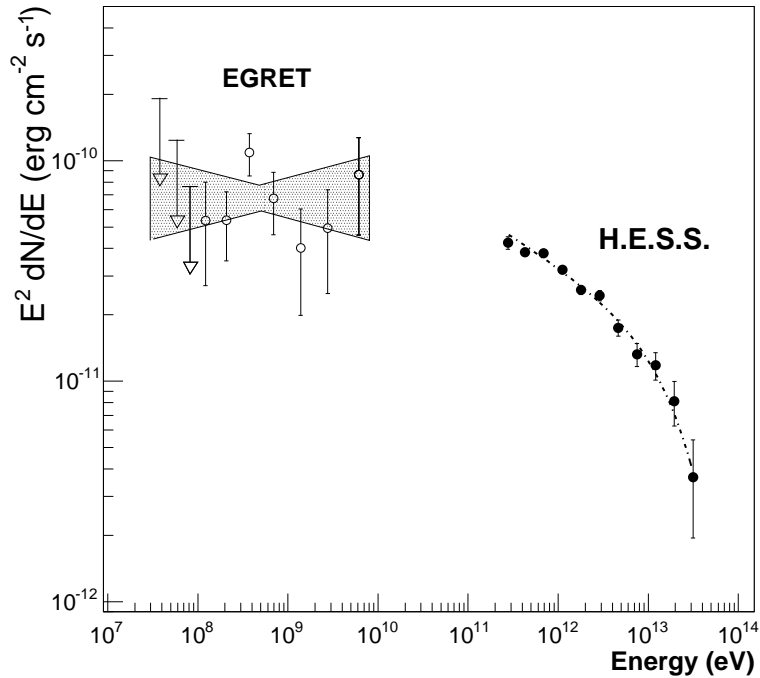,width=10cm,clip=} }
\caption{The photon spectra of HESS\,J1825-137 and the nearby source
3EG J1826-1302 (multiplied by $E^2$) showing the curvature in the spectrum
of HESS\,J1825-137 \cite{Aha06c}.}
\label{fig:6}        
\end{figure}

The above mentioned behaviour is also summarised when we plot the total spectrum of HESS\,J1825-137
as shown in Figure \ref{fig:6}, where we see that this photon spectral index steepens
from $\sim 1.9$ to $\sim 2.6$ as expected for a cooling break. It is important
to measure this cooling break, since we can then determine the average nebular
field strength of the PWN, independently from a comparison of synchrotron
and IC brightnesses, for which we do not have comparative data corresponding
to the same electron energies. There are two problems if we attempt to
``read'' this break energy $E_b$ from the energy spectrum as shown in Figure \ref{fig:6}:
(a) We do not know what the convergent (uncooled) photon spectral index
below the H.E.S.S. range is, although this should be reflected by
the synchrotron photon index ($\sim 1.6$) of the uncooled electrons near the PWN shock.
Therefore, most likely the electron spectral index for this domain is around 2.2.
(b) Klein-Nishina effects
due to dust IR photons tend to steepen spectra with increasing
photon energy as described by \cite{dej95} for such PWN. 
A more accurate procedure would then be either through direct modelling
as done by \cite{Lem06} for HESS\,J1825-137, or, simply through
fitting a two component electron spectrum scattering the CMBR and dust IR
photons and rewriting the expressions for the cooling break directly in terms
of the electron spectral break energy. Fortunately the estimate for the
nebular field strength depends weakly on $E_b$ as discussed below:

The anomalously large size of the unseen SNR forward shock radius is best
met if the pulsar braking index $n$ is closer to 2 than 3, giving a pulsar spindown age
\begin{equation}
T_p=(40\,{\rm kyr})\left(\frac{1}{n-1}\right),
\end{equation}
which is closer to 40 kyr, rather than the canonical 20 kyr.
It is possible to probe the field strength in most of the PWN volume
by solving the expression
\begin{equation}
-\int_{E({\rm max})}^E\frac{dE_e}{(dE_e/dt)_s+(dE_e/dt)_{\rm IC}}=T_p
\end{equation}
for the electron energy $E_e$, where we include both synchrotron and
inverse Compton losses \cite{Blu70}. In this case electrons
injected with the maximum energy $E_e({\rm max})$ at the PWN shock during the earliest
epochs (when the spindown power was a maximum, see Section \ref{sec:3}), move with the outer edge of the PWN.
While losing most energy in the PWN, they must still be able to radiate VHE photons with energy at 
least $\sim 4$ TeV to account for the highest energy
spectral point at a distance of $\sim 1$ degree from the pulsar as shown in Figure 4 of
\cite{Aha06c}. 

For inverse Compton energy losses on the 2.7K CMBR we include a Klein-Nishina suppression factor (relative to the Thomson limit) 
of $2/3$ in the H.E.S.S. range \cite{Aha06c}.
By setting the spindown age equal to the total radiation lifetime of VHE $\gamma$-ray emitting electrons, 
we can write the electron energy in terms of the VHE $\gamma$-ray energy (Eqn \ref{Eeg}) to give the observed spectral break energy
of $E_b\sim 2.5$ TeV \cite{Lem06}, which is observed as a steepening in Figure (\ref{fig:6}):
\begin{equation}
E_b=\frac{(6.2\,{\rm TeV})(n-1)^2}{[1+0.144(B_{\mu{\rm G}})^2]^2}.
\label{egb}
\end{equation}
Thus, for $n=2$ ($T=40$ kyr age) and a $B\sim 2\mu$G field, the VHE $\gamma$-ray break would be around 2.5 TeV, whereas
for $n=3$ (i.e. a $T=20$ kyr age), the required field strength would be $3.9\mu$G.
Whereas \cite{Lem06} found the abovementioned break energy from broad band modelling of HESS\,J1825-137,
GLAST should be able to measure the uncooled spectral index at $\gamma$-ray energies $E_{\gamma}\ll E_b$, which will allow
us to constrain $E_b$ more accurately in future.

\subsection{Conclusion: A particle dominated wind in HESS J1825-137}
With both the spectral break around a TeV and the survival of $\sim 5$ TeV emitting electrons
to the edge of the PWN, it is clear that we require a
magnetic field strength $B<3\mu$G in the extended nebula. Such a low field strength
also supports the relatively low observed X-ray to VHE $\gamma$-ray luminosity. A detailed
treatment of this is however beyond the scope of this paper.

A concern which may be raised from a lay perspective: We know that the field strength
in the ISM is about $3\mu$G or more, then why do we get an apparent field strength below this value?

The total pressure in the PWN is the sum of the magnetic pressure plus particle pressure, and
we can already derive the total particle pressure from the electron spectrum responsible 
for the {\it HESS} signal: \cite{dej07} derived the pulsar pair production multiplicities
from the {\it HESS} data alone, as well as an upper limit by extrapolating the {\it HESS} spectrum along the  
harder uncooled pre-break $e^{+}-e^{-}$ spectrum with an index of $\sim 2$ down to $E_0\sim 1$ GeV
\cite{dej07}. The total energy in electrons in the H.E.S.S. range 
is $\sim 10^{48}$ ergs, but if we take
the total energy down to $E_0$ also into account,
the total energy would be $E_{\pm}\sim 8\times 10^{48}$ ergs.
This would have required a pulsar birth period $2\pi/\Omega_0=P_0<50$ ms to give a total rotational
kinetic energy of $0.5I\Omega_0^2> E_{\pm}$.

The total energy density in leptons is then $U_e\sim (1\,{\rm eV/cm}^3)d_4^{-3}$ for
a PWN radius of $0.5^{\circ}$. 
The accuracy of this number is expected to improve
when GLAST observations are added, which should measure the spectral hardening well below the break
with better accuracy. The energy density in a $B=2\,\mu$G PWN field is  $U_B=(0.1\,{\rm eV/cm}^3)(B/2\,\mu{\rm G})^2$.
Thus, to a first order we find that $U_e\sim 10U_B$, so that
\begin{itemize}
\item The PWN of HESS\,J1825-137 is particle dominated with lepton energy density about 10 times the field
energy density, thus adding this PWN to the \cite{Che04} list of particle dominated winds;
\item The pressure $U_e$ still appears to be significant to press against the ISM medium,
which resulted in the anomalously large PWN as observed today.
\end{itemize}

\section{Vela X -- The prototype for evolutionary studies\label{velax}}

\begin{figure}
\centerline{\psfig{file=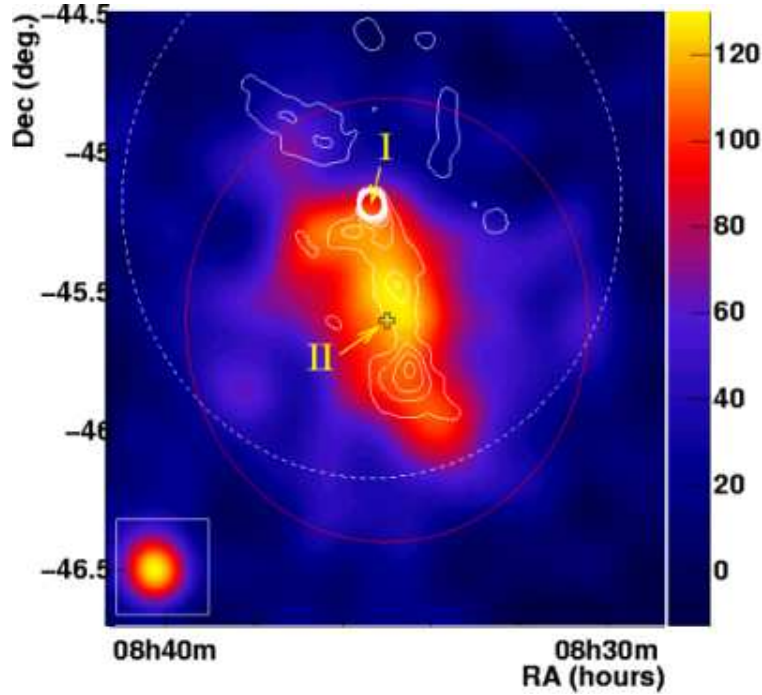,width=10cm,clip=} }
\caption{Gaussian smoothed sky map of region surrounding Vela pulsar,
showing significant emission to the south of the pulsar position,
coincident with an X-ray feature seen by ROSAT PSPC (white contours).
The solid circle represents the H.E.S.S. integration region
for the spectral measurement, while the dashed circle represents
the field of view for the ROSAT observations. 
(From \cite{Aha06a}; see also this paper for more details).}
\label{fig:7}        
\end{figure}

Vela X, the bright flat spectrum radio component of the Vela SNR served as the prototype
PWN for evolutionary studies \cite{Wei80}. The offset of the radio
nebula to the south of the pulsar had Astronomers doubting if this association is real,
until ROSAT discovered a cocoon of X-ray emission, also extending south of the pulsar, and
aligned with a bright radio filament in Vela X \cite{Mil95}. The length of both features
to the south is $\sim 45'$. More revealing was that even though this is one of
the brightest polarised radio filaments, the degree of polarisation is low ($\sim 15\%$ to 20\%).
The reason for this is the presence of thermal material mixed into the plasma of
highly relativistic particles \cite{Mil95}. A natural explanation for this two-fluid mixing and offset to
the south was offered by \cite{Blo01} for Vela X: The reverse shock returning from
a denser ISM offsets the PWN to the south while forcing this two-fluid mixing.
Further evidence for this mixing came from the combined ASCA/ROSAT analysis of the bright cocoon
(radio filament) region by \cite{Mar97} which shows evidence of mixing
of a non-thermal component with photon index of $\sim 2$ with a thermal component.
Analyses by \cite{Hor06} with new ASCA results, as well as {\it XMM-Newton}
observations of the cocoon of Vela X also confirmed this two-component spectrum
\cite{LaM06}.

\subsection{H.E.S.S. detection of the Vela X ``cocoon'' -- radio \& X-ray correlation}
The High Energy Stereoscopic System of telescopes observed the Vela region and discovered 
a structure resembling the Vela X cocoon in X-rays \cite{Aha06a} as shown
in Figure (\ref{fig:7}). The relative sizes however differ significantly:
Whereas the size of the X-ray cocoon is $45\times 12$ arcmin$^2$, the corresponding 
VHE cocoon size is $58\times 43$ arcmin$^2$.

\begin{figure}
\centerline{\psfig{file=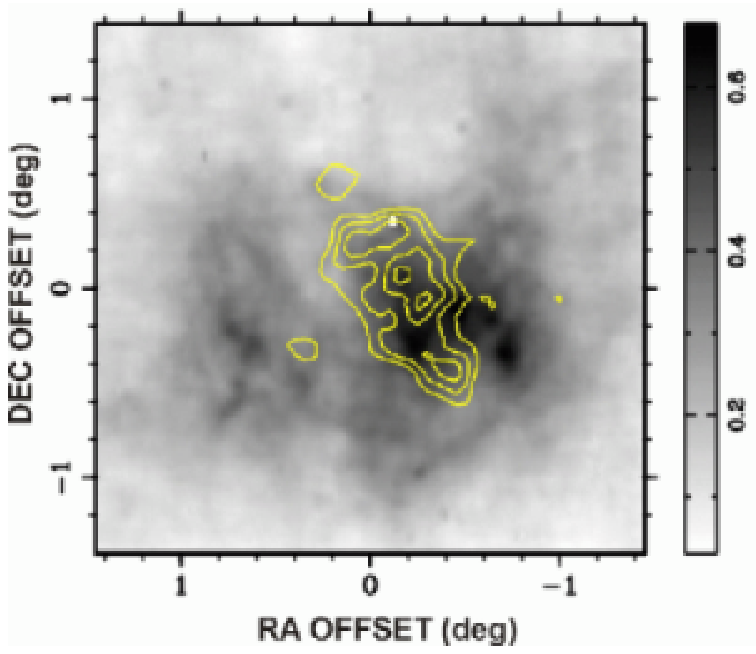,width=10cm,clip=} }
\caption{Destriped radio image of Vela X at 8.4 GHz in greyscale \cite{Hal04}
with the {\it HESS} map in Figure (\ref{fig:7}) converted to contours, overlayed in yellow. 
The pulsar position is marked by a white circle.
The origin (zero point) of this image is (RA,DEC) =(129.02$^{\circ}$,-45.54$^{\circ}$) (J2000),
which is different from the origin defined by \cite{Hal04}.}
\label{fig:8}        
\end{figure}

To complete the multiwavelength comparison, we also compare the VHE detection with
the radio map: Figure (\ref{fig:8}) shows an overlay of the {\it HESS} $\gamma$-ray contours on the 8.4 GHz radio map
of Vela X \cite{Mil95,Hal04}, 
showing that the $\gamma$-ray map does overlap with the bright radio filament,
which in turn overlaps with the X-ray cocoon as remarked by \cite{Mil95}. The latter author
made the following observation based on the \cite{Rey88} model: 
The radio filaments are the result of Rayleigh Taylor instabilities in the SNR expansion.
We conclude further that these 
filaments were also offset to the south of the pulsar by the early reverse shock.
If the $\gamma$-ray signal is then due to hadronic interactions with this thermal gas,
we would also expect to see a correlation between the $\gamma$-ray and filamentary structures.

The approximate full size of the Vela X PWN is $\sim 3^{\circ}$ in RA and $\sim 2^{\circ}$
in DEC as seen from a $8^{\circ}\times 8^{\circ}$ HartRAO radio map of the Vela region at 2.3 GHz by Jonas (2006, personal
communication). Note that the status of VHE $\gamma$-radiation from Vela X as a whole was not discussed by \cite{Aha06a},
although this total flux cannot be much larger than that from the VHE cocoon itself.
We can then summarize the morphological multiwavelength properties of the cocoon detection as follows:

\begin{itemize}
\item The volume of the VHE $\gamma$-ray cocoon is $\sim 5$\% relative to the total volume of Vela X;
\item The cocoon contains both a non-thermal and thermal component, which is indicative of
a reverse shock mixing gases of adiabatic indices $\gamma=4/3$ and 5/3 during the crushing phase;
\item The offset of the cocoon to the south of the pulsar is explained by the reverse shock
crushing the PWN to the south (see next Section); 
\item The position angle (P.A.) of the VHE $\gamma$-ray cocoon ($41\pm 7$ degrees, measured from north through east) 
is similar to that of the X-ray cocoon orientation (see next Section).
\item The X-ray cocoon overlaps in position with a bright radio filament, although
there are other similarly bright radio filaments further to the west without any X-ray or VHE
$\gamma$-ray counterparts;
\item Deeper, but wide FoV VHE observations of the entire Vela X structure
shown in Figure (\ref{fig:8}) should indicate if there are $\gamma$-rays with lower surface brightness 
compared to the bright cocoon region. 
\end{itemize}

\subsection{Constraints on the cocoon field strength from the upper synchrotron cutoff energy}


Since the cocoon has been shifted to the SW of the pulsar by the reverse shock,
the highest energy electrons mixed in the thermal gas in the cocoon provide
a powerful diagnostic of the age since shift and associated field strength.
\cite{Mar97} measured a power law component up to at least 7 keV
from the cocoon area. \cite{Hor06} and \cite{LaM06} also confirmed
this two-component composition with {\it ASCA} and {\it XMM-Newton} observations respectively. 

This means that ultrarelativistic electrons and thermal gas were mixed during the
reverse shock crushing phase and with the offset PWN, the relativistic component is
removed from its pulsar source, so that the upper spectral cutoff energy moves
down in energy with time, without any source of replenishment. From Fig. 3 of \cite{Hor06}, it is clear
that this cutoff is currently $\sim 10$ keV. The electrons radiating at this cutoff
are not replenished by the Vela pulsar, since they have already been removed from the pulsar
over a time interval $T-T_c$, which represents the time between the southward shift of the PWN from the pulsar 
(at time $T_c$) and the present time $T$. These 10 keV emitting electrons were also the highest energy electrons
accelerated by the PWN shock at the epoch $T_c$ when the reverse shock started to crush the PWN. 
For such high energy electrons, IC scattering
would be in the extreme Klein-Nishina limit, so that we only consider synchrotron
losses, giving the time interval between the time of crushing and the present time of
\begin{equation}
T-T_c=(2.3\,{\rm kyr})\left(\frac{3\,\mu{\rm G}}{B}\right)^{3/2}\left(\frac{10\,{\rm keV}}{E_X({\rm max})}\right)^{1/2}.
\label{ttc_vela}
\end{equation}
For Vela X \cite{Blo01} calculated $T_c\sim 3$ kyr, whereas 2-D time dependent
MHD simulations for the Vela SNR shows that the reverse shock was expected to reach the
pulsar position around 5 kyr after the birth of the pulsar. Thus, a field strength around
$3\mu$G (or smaller) in the cocoon of Vela X is required to allow 10 keV synchrotron emitting electrons
to survive between the time of crossing of the reverse shock and the present epoch. If the
field strength was $10\mu$G or larger (as required by \cite{Hor06} these
electrons had to be shifted within 400 yr, which is unlikely to be achieved given any realistic reverse
shock parameters. 

\subsection{Diffusion of VHE particles from the cocoon}
\cite{Hor06} considered the problem of diffusion of X-ray synchrotron emitting 
electrons away from the X-ray cocoon, stating that a high field strength is required
to contain the ultrarelativistic electrons in the $45\times 12$ arcmin$^2$ cocoon.
Assuming that the cocoon is still expanding under its own pressure, a perpendicular field
component is expected to be maintained by virtue of Eqn (\ref{toroidal}), so that Eqn (\ref{f})
with $f\ll 1$ for $\Omega\tau \ll 1$ or $\Omega\tau\gg 1$ is expected to hold, which
protects the integrity of this PWN against diffusive escape. Replacing the time $T-T_c$
with Eqn (\ref{ttc_vela}) in the diffusion equation, the electron energy cancels,
so that the angular spread due to diffusion at a distance of $d=0.3$ kpc can be
written as 
\begin{equation}
\theta_{\rm diff}=0.5^{\circ}\left(\frac{f}{0.1}\right)^{1/2}\left(\frac{3\,\mu{\rm G}}{B}\right)^{3/2}
\end{equation}
With a minimum X-ray cocoon dimension of 0.2 degrees, it is clear that we have to set $f<0.02$
for $B=3.3\mu$G, which places a restriction on the scattering parameter $\Omega\tau$ 
based on hard-sphere scattering \cite{Ste88}.

\subsection{No ``missing'' leptonic component in Vela X}
\cite{Hor06} suggested that there is a missing leptonic component in Vela X.
Whereas this is true for the cocoon, the actual volume of the Vela PWN (called ``Vela X'') is about
20 times larger than the size of the cocoon as seen in VHE and to get the
total energy in leptons, we have to take the bolometric spectrum from
the total Vela X, which is one of the brightest radio nebulae in the sky. 
Using the radio spectrum of Vela X, \cite{dej07} found that the total lepton energy in the radio nebula
is $W_e=6.2\times 10^{47}$ ergs (for $B=10\mu$G) or $3.8\times 10^{48}$ ergs ($B=3\mu$G), giving
respective conversion efficiencies of $W_e/(0.5I\Omega_0^2)=5\%$ and 30\% for a 
birth period of 40 ms \cite{van01}. 
Thus, there does not appear to be a ``missing'' leptonic component and it is
clear is that most lepton energy has been processed in the low energy leptonic domain. 
However, there are also observational lower limits to the Vela X averaged field strength:
Using EGRET upper limits, \cite{dej96b} have shown that we can already constrain the volume averaged field strength 
to $<\!B\!>\,>\,(4\,\mu{\rm G})(\nu_b/10^{11}\,{\rm Hz})^{0.4}$ where $\nu_b$ is the unknown radio spectral break frequency. GLAST/LAT observations should either detect this radio counterpart of Vela X, or, provide more stringent lower limits on $<B>$. This also calls for a separate study on variations in $B$: How does $B$ in the cocoon differ relative to $<B>$, given the presence of filaments in the PWN as well as the filling factor question?  

\subsection{The H.E.S.S. signal: hadrons or leptons?}
In the previous two sections we have shown the field strength in the cocoon must be relatively
low for X-ray emitting electrons to survive at the southern tip of the cocoon, which
would argue for an IC origin. This also
implies a limit on the scattering parameter $\Omega\tau$ for containment against diffusion
through a weak perpendicular field. We have no theory to predict this number, 
but future research on turbulence theory may be able to make some predictions.

Another potential problem with a hadronic interpretation is the following:
With the high required ion energy budget of $10^{48}$ ergs (iron) to $10^{49}$ ergs (protons)
in the VHE cocoon, \cite{Hor06} invoked the early epoch of pulsar
output to account for the observed $\gamma$-ray flux via hadronic interactions. 
However, protons ejected during such early epochs (and convected
by the pulsar wind) should fill the total (old) radio emitting Vela X PWN and not just the smaller (and younger) cocoon.
Thus, the total energy budget in Vela X implied by the {\it HESS} detection will then be (to a first order)
20 times larger than calculated for the cocoon:
$\sim 2\times 10^{49}$ ergs (for iron) to $2\times 10^{50}$ ergs (for protons). 
Furthermore, \cite{van01} estimated a birth period of $P_0\sim 40$ ms
for the Vela pulsar to account for the observed classical ratio of PWN radius to SNR radius of 0.25. 
The total integrated kinetic energy provided by the pulsar since birth is then $0.5I\Omega_0^2=1.2\times 10^{49}$ ergs,
which means that we may have a conversion efficiency of $>100\%$ of spindown power to ions in Vela X.
Thus, the VHE signal is more likely to be of leptonic than hadronic origin.

\subsection{The VHE $\gamma$-ray spectral break in the Vela X cocoon}
\begin{figure}
\centerline{\psfig{file=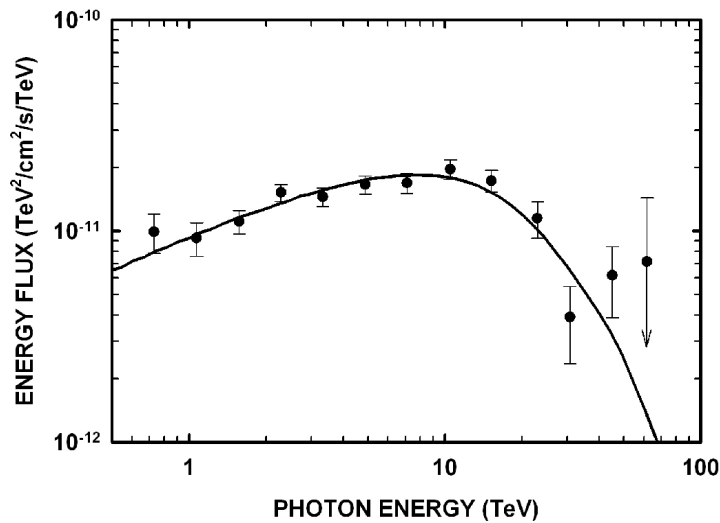,width=10cm,clip=} }
\caption{Energy spectrum of the Vela X cocoon as measured by H.E.S.S. \cite{Aha06a}, 
with integration area shown in Figure \ref{fig:7}. The solid line represents a fit assuming
an inverse Compton origin as specified in the text.}
\label{fig:12}        
\end{figure}

If we assume that the H.E.S.S. signal is due to IC scattering on the CMBR in a relatively weak field of $B\sim 3\mu$G,
then we should be able to predict the cooling break as that energy where radiation losses
become comparable to the age of the PWN/SNR. This may then explain the spectral break in Fig. (\ref{fig:12}):
In this case Eqn (\ref{egb}) would predict a VHE $\gamma$-ray spectral
break energy of $E_b\sim 12$ TeV for such a field strength, given a total age of 11 kyr. Note that
this break energy $E_b$ evolves downward in energy with time, and if we assume that the cocoon field strength
did not change significantly over the past few thousand years, the break energy is predicted to be
\begin{equation}
E_b=\frac{(15\,{\rm TeV})}{[1+0.144(B/{3\,{\mu{\rm G}}})^2]^2}\left(\frac{11\,{\rm kyr}}{T}\right)^2.
\label{egb_vela}
\end{equation}

Note also that Fig. \ref{fig:12} represents \emph{the first detection of a spectral maximum in VHE $\gamma$-ray Astronomy},
which allows this clear measurement of a spectral break energy.

\section{Summary\label{sum}}
In this review paper of pulsar wind nebulae (as seen by H.E.S.S.), we attempted to interpret the
observed properties within a unified framework: We first reviewed the known properties of PWN
as derived over many years from radio, IR, optical and X-ray observations. We then introduced new properties
which are expected to make PWN bright high-energy to VHE $\gamma$-ray sources: For example, if the
energy density in the soft radiation field (acting as target for IC scattering) dominates the energy
density of the magnetic field, electron energy losses would be dictated by the IC rather than the synchrotron
process. This principle is demonstrated when we compare the general properties of H.E.S.S. PWN with the Crab Nebula:
The Crab Nebula is an efficient synchrotron radiator but inefficient $\gamma$-ray emitter 
(i.e. a high ratio of optical/X-ray energy flux to VHE $\gamma$-ray flux) as a result of the relatively large magnetic
field energy density, but for most other H.E.S.S. PWN, the ratio of synchrotron to IC energy fluxes are comparable
to, or even less than unity. This then hints at a relatively small magnetic energy density. 
Furthermore, in such cases we may also find that the energy density in relativistic
electrons dominates the magnetic energy density, leading to the description of ``{\bf particle dominated winds}'',
as opposed to the Crab Nebula which is known to be in equipartition.

We then reviewed the $\gamma$-ray lifetimes of PWN by considering electrons losing energy due to
both synchrotron and inverse Compton radiation in the expanding post-shocked flow. 
As the PWN expands well beyond its X-ray phase (this X-ray phase terminates when the
overall field strength becomes too small as a result of expansion), IC losses on the CMBR is then expected to dominate, 
in which case the terminating lifetime would converge to a value of 100 kyr. 
{\bf Thus, if we observe a PWN at an energy near 1 TeV, the liftime of the PWN is expected to be $\le 100$ kyr.}
This discussion naturally led to the concept of particle spectral steepening as a function
of increasing radius, but only as long as either synchrotron or IC radiation dominates the electron energy loss process. 

The concept of dispersion in a PWN was also discussed: Is the observed
size of a PWN mostly due to post-shocked convective (pulsar wind) flow, or, would diffusion
dominate the process? At first glance (from this review paper) it seems as if the ordered magnetic field in a PWN would
inhibit dispersion due to diffusion. The result of this is that the expanding PWN "bubble"
contains its radiating charged particles, but only as long as the magnetic field direction
maintains its perpendicular direction relative to the radial direction. This question is relevant if we want
to calculate the luminosity of a PWN as a function of time.

This review paper then concluded with a discussion of two important H.E.S.S. sources, where
aspects such as (i) size, (ii) spectral energy distribution (SED) maximum,
(iii) the offset of the center of gravity (of the VHE emission) relative to the pulsar position as 
a result of SNR expansion into an inhomogeneous interstellar medium,
(iv) energy dependent morphology (i.e. the effect of spectral steepening in the radial direction)
and (v) the observed VHE spectral break were discussed (if interpreted as the SED maximum). 
For both these sources it appears
as if IC losses dominate over synchrotron losses, which has the advantage that the
age of the PWN can be relatively accurately determined from the observed VHE spectral break: Assuming this,
we showed that we indeed get consistent ages for both {\it HESS} sources.
\section{Acknowledgements}
The first author acknowledges support from the South African Department of
Science \& Technology and National Research Foundation Research Chair: Astrophysics \& Space Science.
Support from the GDRI-GREAT French, German, South African \& Namibian multinational funding source is also acknowledged.
The authors would like to thank members of the Supernova Remnant, Pulsar and Plerion working group of the
H.E.S.S. collaboration for useful discussions.

%
%

%
%



\printindex
\end{document}